\documentclass[10pt,conference]{IEEEtran}
\usepackage[T1]{fontenc} %
\usepackage{cite}
\usepackage{amsmath,amssymb,amsfonts}
\usepackage{algorithmic}
\usepackage{graphicx}
\usepackage{textcomp}
\usepackage[dvipsnames]{xcolor}
\usepackage[all]{nowidow}
\usepackage[caption=false]{subfig}
\usepackage{balance}
\usepackage{hyperref}
\usepackage[capitalise,noabbrev]{cleveref}
\usepackage{tikz}  %
\usepackage{float}
\usepackage{nowidow}

\usepackage[disable]{todonotes}
\newcommand*\circled[1]{\tikz[baseline=(char.base)]{
\node[shape=circle,draw,inner sep=0.75pt,scale=0.8,font=\fontfamily{phv}\selectfont] (char) {#1};}%
}

\newcommand{\name}{\text{\textsc{Fusionize}}}

\newcommand{\fancytt}[1]{\texttt{%
\hyphenchar\font=`\- %
\hyphenpenalty=10000 %
\exhyphenpenalty=-50 %
#1\ignorespaces
}}

\newcommand{\textsfcmu}[1]{%
{\fontfamily{cmss}\selectfont #1}\ignorespaces
}

\def\BibTeX{{\rm B\kern-.05em{\sc i\kern-.025em b}\kern-.08em
    T\kern-.1667em\lower.7ex\hbox{E}\kern-.125emX}}
\begin{document}

\title{\name{}: Improving Serverless Application Performance through Feedback-Driven Function Fusion}

\author{\IEEEauthorblockN{Trever Schirmer\IEEEauthorrefmark{1}, Joel Scheuner\IEEEauthorrefmark{2}, Tobias Pfandzelter\IEEEauthorrefmark{1}, David Bermbach\IEEEauthorrefmark{1}}
    \IEEEauthorblockA{\IEEEauthorrefmark{1}\textit{TU Berlin \& ECDF, Mobile Cloud Computing Research Group}\\
        \{ts,tp,db\}@mcc.tu-berlin.de}
    \IEEEauthorblockA{\IEEEauthorrefmark{2}\textit{Chalmers | University of Gothenburg}\\
        scheuner@chalmers.se}
}

\maketitle

\begin{abstract}
    Serverless computing increases developer productivity by removing operational concerns such as managing hardware or software runtimes.
    Developers, however, still need to partition their application into functions, which can be error-prone and adds complexity:
    Using a small function size where only the smallest logical unit of an application is inside a function maximizes flexibility and reusability.
    Yet, having small functions leads to invocation overheads, additional cold starts, and may increase cost due to double billing during synchronous invocations.
    In this paper we present \name{}, a framework that removes these concerns from developers by automatically fusing the application code into a multi-function orchestration with varying function size.
    Developers only need to write the application code following a lightweight programming model and do not need to worry how the application is turned into functions.
    Our framework automatically fuses different parts of the application into functions and manages their interactions.
    Leveraging monitoring data, the framework optimizes the distribution of application parts to functions to optimize deployment goals such as end-to-end latency and cost.
    Using two example applications, we show that \name{} can automatically and iteratively improve the deployment artifacts of the application.
\end{abstract}


\section{Introduction}
\label{sec:introduction}

In recent years, serverless computing as the latest incarnation of cloud computing has become more and more popular~\cite{Hendrickson2016-pw,paper_bermbach_cloud_engineering}.
In serverless computing, developers rely on managed cloud platform services as application building blocks which are connected through application-specific glue code.
Usually, this glue code is written as sequences of small, stateless functions~\cite{paper_grambow_befaas,jia2021nightcore} -- often invoked in an event-driven model -- running on top of Function-as-a-Service (FaaS) platforms such as AWS Lambda\footnote{\url{https://aws.amazon.com/lambda}} or Google Cloud Functions\footnote{\url{https://cloud.google.com/functions}}.

\begin{figure}
    \includegraphics[width=\linewidth]{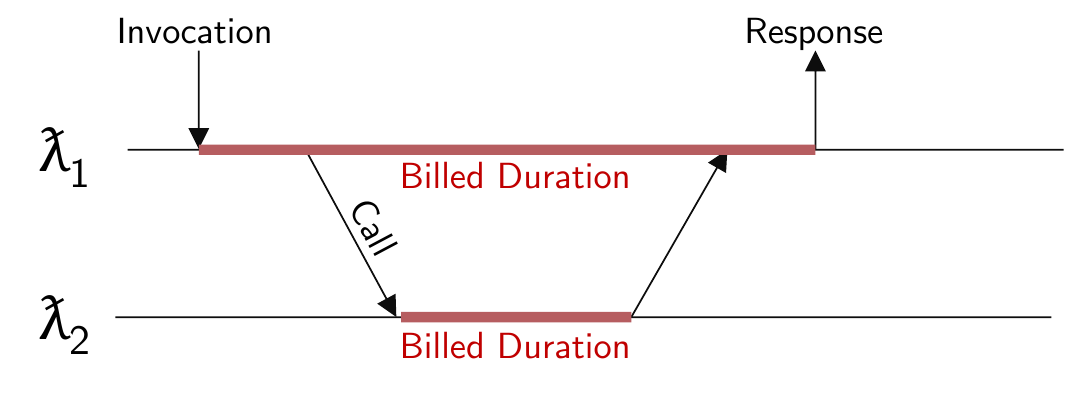}
    \caption{Synchronous invocations of functions can lead to double billing in the duration-based billing model of FaaS: While ${\lambda}_{1}$ calls ${\lambda}_{2}$, both functions incur costs.}
    \label{fig:doublespending}
\end{figure}

While some work has been devoted to sizing of functions, e.g.,~\cite{Akhtar_2020,Elgamal_2018,Czentye_2019}, the question of an optimal function size has not been solved yet:
From a software design perspective, functions should be small and cover only an isolated piece of functionality to maximize flexibility and reusability.
From a performance perspective, however, fine-grained functions lead to increased invocation overheads and cold starts~\cite{paper_bermbach_faas_coldstarts}, i.e., having only a single large function would be optimal.
From a cost perspective, functions should prevent waiting for other functions to finish, since this leads to double billing as shown in \cref{fig:doublespending}~\cite{Baldini_2017_Trilemma}.
Finally, from a platform perspective, functions need to comply with platform limits regarding resource usage or execution duration which are platform-dependent and may change over time.

In this paper, we propose to close this gap by differentiating between the developer-defined software design artifact (\emph{tasks}) and the deployment artifact (\emph{functions}) through a system called \name{}:
During software design, developers follow a fine-granular programming model, composing tasks to build their application.
Then, during the deployment process, the \name{} framework uses function fusion to transform the software design artifact into the actual deployment artifact.
Starting from a basic deployment artifact, \name{} deploys the artifact and uses monitoring data to gradually transform the deployed artifact towards a more efficient version: an artifact with an optimal cost-performance profile, adhering to platform limits.
The \name{} architecture can be used by both cloud users and providers.
For this, we make the following contributions:

\begin{itemize}
    \item We propose \name{}, a feedback-driven deployment framework that iteratively adapts the deployment of FaaS applications at runtime using an optimization strategy (\cref{sec:system}).
    \item We discuss a first heuristic for such an optimization strategy (\cref{sec:heuristics}).
    \item We implement \name{} as an open-source prototype and present the results of extensive experimentation with two example applications (\cref{sec:evaluation}).
    \item We discuss the limitations of our approach and derive avenues for future work (\cref{sec:discussion}).
\end{itemize}
\section{Background}
\label{sec:background}
In this section, we give a brief overview of FaaS as a paradigm, discuss influence factors on performance and cost, and finally function fusion as a concept.

FaaS is a popular paradigm for cloud software development and deployment in which developers write their source code in the form of small, stateless functions\footnote{In the later sections of the paper, we will use the terms ``tasks'' to refer to the developer-defined functions and the term ``function'' to refer to the resulting deployment artifact.} and leave all operations aspects to the cloud provider.
These functions are usually written in high-level programming languages such as Node.js\cite{9251194}.
At runtime, the cloud provider routes incoming requests or events to an instance of the corresponding function -- either to a newly created one thus incurring the so-called cold start latency~\cite{paper_bermbach_faas_coldstarts,daw2020xanadu} or reusing an existing one.
In practice, typical FaaS applications are compositions of short-lived, transient, often highly concurrent function instances~\cite{paper_bermbach_cloud_engineering,paper_bermbach_faas_coldstarts}.

With current FaaS providers, developers have to manually choose the amount of resources (i.e., memory or disk space) they wish to provision per function instance.
Overall cost is then determined based on the resulting resource price multiplied with the execution duration of functions.
Furthermore, there is usually additional cost per request.
Depending on the composition (orchestration vs. choreography~\cite{paper_bermbach_faas_coldstarts,Baldini_2017_Trilemma}), FaaS applications often suffer from double billing when one function makes a blocking call to another one~\cite{Baldini_2017_Trilemma}.
For optimizing costs, it is hence crucial to avoid double billing and the provisioning of unused resources.

Performance of FaaS functions again depends on the amount of provisioned resources, on cold starts, and the number of requests.
Of these, the amount of resources has only a small effect whereas the performance costs for function calls are significant.
For instance, in AWS function calls are routed through the API Gateway and in Apache OpenWhisk requests are usually queued in Apache Kafka before execution.
On top of this, cold start latency depends on programming language and other parameters but is generally in the range of about one second.
Furthermore, increasing load can cause cascading cold starts in a function composition~\cite{paper_bermbach_faas_coldstarts,daw2020xanadu}.
From a performance perspective, it is hence crucial to have as few invocations as possible which are not only costly but also each carry the risk of encountering a cold start.
The mechanism to address this has been referred to as function fusion~\cite{Elgamal_2018}: Two or more functions are rewritten (and redeployed) as one.
\section{Automatically Adapting Functions at Runtime}
\label{sec:system}

In this section, we give an overview of \name{} and describe how it can automatically adapt FaaS functions at runtime.
We start by describing key terms and definitions (\cref{subsec:terms}) before discussing \name{} and its components (\cref{subsec:approach}) and the resulting programming model (\cref{subsec:progmodel}).

\subsection{Terms and Definitions\label{subsec:terms}}
As already introduced earlier, we use the term \textbf{\emph{task}} to refer to the function written by the developer and the term \textbf{\emph{function}} to refer to the executable deployment artifact.
Each function contains a single \textbf{\emph{fusion group}}, i.e., one or more tasks that are executed as part of that specific function.
An application comprises one or more (FaaS) functions (and hence at least as many tasks and fusion groups).
While an application is deployed, the assignment from tasks to fusion groups can change dynamically.
We refer to a specific set of fusion groups, i.e., one possible deployment configuration of all tasks in an application, as \textbf{\emph{fusion setup}}.
In this paper, we use a short notation to describe fusion setups in the text and in figures:
Tasks that are in the same fusion group are put in parentheses, separated with commas; the fusion setup separates fusion groups with hyphens.
As order only matters for execution but not for deployment, we order all task names alphabetically.
For example, in the fusion setup \texttt{(A,B)-(C)}, the tasks \texttt{A} and \texttt{B} are in the same fusion group, and task \texttt{C} is in its own group.

\subsection{The \name{} Approach\label{subsec:approach}}

\begin{figure*}
    \includegraphics[width=\textwidth]{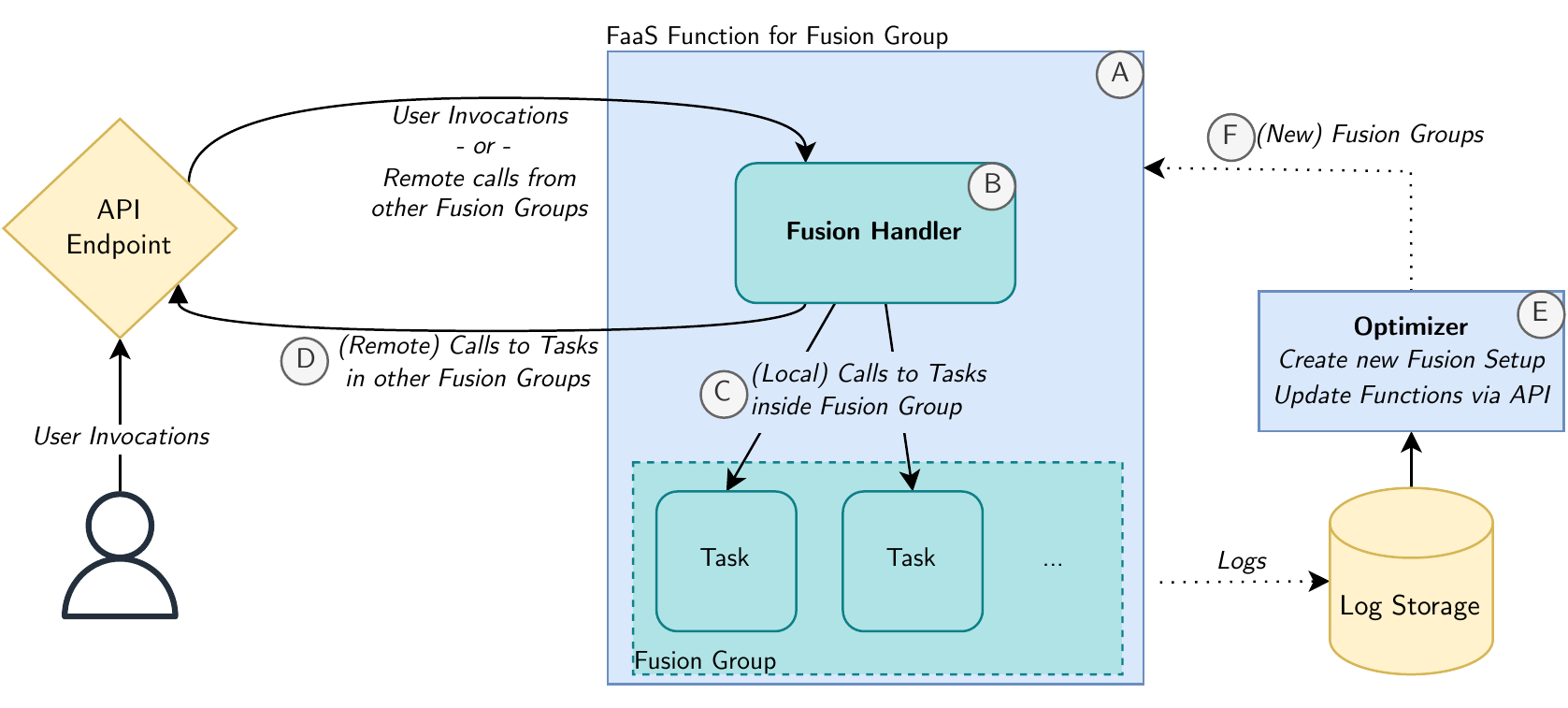}
    \caption{Overview of the Architecture of \name{}: Within a FaaS function, the fusion handler relays invocations to different tasks in its fusion group. Tasks can call other tasks, which are executed locally if the task is in the same fusion group, or remotely in the function of another fusion group. The Optimizer regularly analyzes function logs to update the fusion setup and functions.}
    \label{fig:approach:architecture}
\end{figure*}

The optimal fusion setup depends not only on developer preferences (regarding cost and performance goals which may be in conflict) but also on runtime effects.
A key design goal for \name{} was hence to implement a feedback-driven, autonomous deployment framework that collects and uses FaaS monitoring data to iteratively fuse tasks to optimize the fusion setup.

As we show in a high-level overview of our approach in \cref{fig:approach:architecture}, \name{} has two main components: the \emph{Fusion Handler} and the \emph{Optimizer}.
In addition, there are also the Log Storage, responsible for storing monitoring data, and the API Endpoint which is the FaaS platform specific service accepting requests to FaaS functions, e.g., API Gateway on AWS.

The Fusion Handler is responsible for request dispatching from and to tasks.
As there is one function per fusion group (\circled{A} in \cref{fig:approach:architecture}), the Fusion Handler is implemented as a distributed component which is co-deployed with every function similar to the choreography middleware in~\cite{paper_bermbach_faas_coldstarts}.
From the FaaS provider perspective, the Fusion Handler is the endpoint for incoming calls to that function (\circled{B}).
Internally, the Fusion Handler calls the requested tasks either locally (\circled{C}) or remotely (\circled{D}).

We also collect monitoring data from each function.
This data includes the execution durations of each of the tasks, memory usage of the function, and which tasks are called from each of the tasks in the fusion group.
In our prototype, we rely on the logging features of the FaaS platform to collect said monitoring data.
In case of a FaaS platform not supporting monitoring or logging, the Fusion Handler could easily be extended to also measure the desired metrics and send them to an external, self-implemented monitoring application as the last activity before returning after a call.

The Optimizer retrieves monitoring data from the Log Storage (\circled{E}), automatically derives the call graph of the application, and annotates it with execution information, e.g., latency values.
In a second step, it uses an extensible optimization strategy module (we describe a first heuristic for this in \cref{sec:heuristics}) to derive an improved fusion setup and triggers a redeployment of all functions that have been changed (\circled{F}).
We propose to use an adapted version of the continuous sampling plan (CSP-1)~\cite{dodge1943sampling,bermbach2011extendable} to decide when to run the Optimizer:
The original algorithm uses the quality of previous items in a factory production process to decide when to run the next quality inspection.
In the adapted version, we propose to compare cost and performance metrics of the monitoring snapshots considered during the previous and the current Optimizer run.
This way, the Optimizer will be run frequently for a newly deployed application but will still from time to time check performance and cost of ``older'' applications.
Applications can hence adapt in the presence of cost or performance changes resulting from external factors such as load profiles or changes to platform limits.

Within the Optimizer, the ``best'' fusion setup can be determined in various ways.
For instance, we could choose the fusion setup with the lowest median latency, which will be the overall fastest fusion group, or we could choose the fusion setup with the lowest 99 percentiles latency, which will be the fusion setup with the fewest cold starts.
Similarly, we could consider various cost metrics.
As part of the optimization strategy, application developers should here assign weights to different optimization goals.

\subsection{The \name{} Programming Model\label{subsec:progmodel}}
From a FaaS developer perspective, the programming model is similar to standard FaaS programming.
This is the case since the Fusion Handler acts as a wrapper, simply exposing the interface of the respective task(s).
The key difference is for outgoing requests:
Tasks here do not make direct calls to the target FaaS function but rather request a call to the target task from the Fusion Handler.
Overall, this means that existing FaaS applications can easily be migrated into the \name{} programming model.

\section{A Heuristic for Optimizing FaaS Functions}
\label{sec:heuristics}
In this section, we introduce a heuristic which improves the fusion setup of FaaS applications using \name{}.
Please note that the focus of this work is on the systems aspects of \name{} and not on the optimization strategy, which we leave to future work.
Here, we propose a set of rules which will allow \name{} to create a \emph{good} (but not necessarily the \emph{best}) fusion setup.

Fusing tasks can have various effects on performance and cost.
Consider the example of a setup in which one task calls another task and has to synchronously wait for the result before continuing.
If the called task is in another fusion group this will lead to double billing.
Thus, those two tasks might benefit from being in one fusion group.
A task calling another task asynchronously does not lead to double billing.
However, the problem of cold starts and call overhead still remains.

\begin{figure}[t]
    \centering
    \includegraphics[width=0.9\columnwidth]{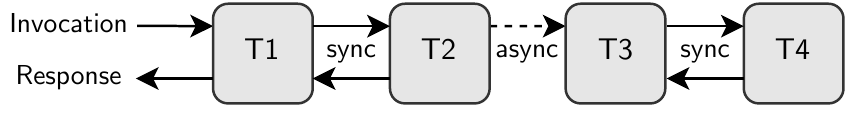}
    \caption{Example call graph that could benefit from function fusion: Task \texttt{T1} and \texttt{T2} need to finish before a response can be sent, and could thus be fused to reduce costs. \texttt{T3} is only called asynchronously, and should thus be moved to another fusion group.}
    \label{fig:approach:example}
\end{figure}

Another example is shown in \cref{fig:approach:example}: \texttt{T1} receives a request, makes a synchronous call to \texttt{T2}, which makes an asynchronous call to \texttt{T3} (e.g., for logging purposes) which finally calls \texttt{T4} synchronously.
\texttt{T1} and \texttt{T2} would benefit from being in one fusion group to avoid double billing and cascading cold starts~\cite{daw2020xanadu}.
Within one fusion group, however, it does not matter whether a task is called synchronously or asynchronously (aside from concurrency aspects) since the function execution has to wait until all its tasks have completed execution.
This means that \texttt{T2} and \texttt{T3} would benefit from being in two different fusion groups as only \texttt{T1}-\texttt{T2} is on the critical path from the application perspective, i.e., also having \texttt{T3} in that fusion group would unnecessarily increase the call latency for the client or function invoking \texttt{T1}, since the function needs to wait until every local task is finished before it can return.

Furthermore, not only tasks that call each other might benefit from being in the same fusion group.
For example, if a task \texttt{A} asynchronously calls either task \texttt{B} or \texttt{C} (and never both) at a fixed rate that can be managed by a single function instance, the amount of cold starts might be reduced if both tasks \texttt{B} and \texttt{C} are handled by the same function instance.
Sometimes, it may also be necessary to split tasks into different fusion groups so that the overall function complies with platform limitations such as maximum execution duration or available disk space.
Depending on this, our tasks \texttt{A}, \texttt{B}, and \texttt{C} might end up either as one fusion group or as fusion groups \texttt{(A)-(B,C)}.

Since there is no difference between calling a local task synchronously or asynchronously, overall request response latency of the original function can be reduced by keeping all long-running asynchronous tasks in separate remote functions.
The total billed duration can be decreased by not waiting on remote function calls (which would lead to double billing) and by moving computationally intensive tasks to their own function so that multiple complex tasks do not block themselves.
\name{} can therefore be especially effective if it can move computationally intensive tasks to their own fusion group so that they do not block the critical path.
There is also the special case where a task uses up all available resources of a function, e.g., by completely filling up the available disk space.
In this case, invocations can only complete if they are placed in their own fusion group.

\begin{figure}[t]
    \centering
    \includegraphics[width=0.45\textwidth]{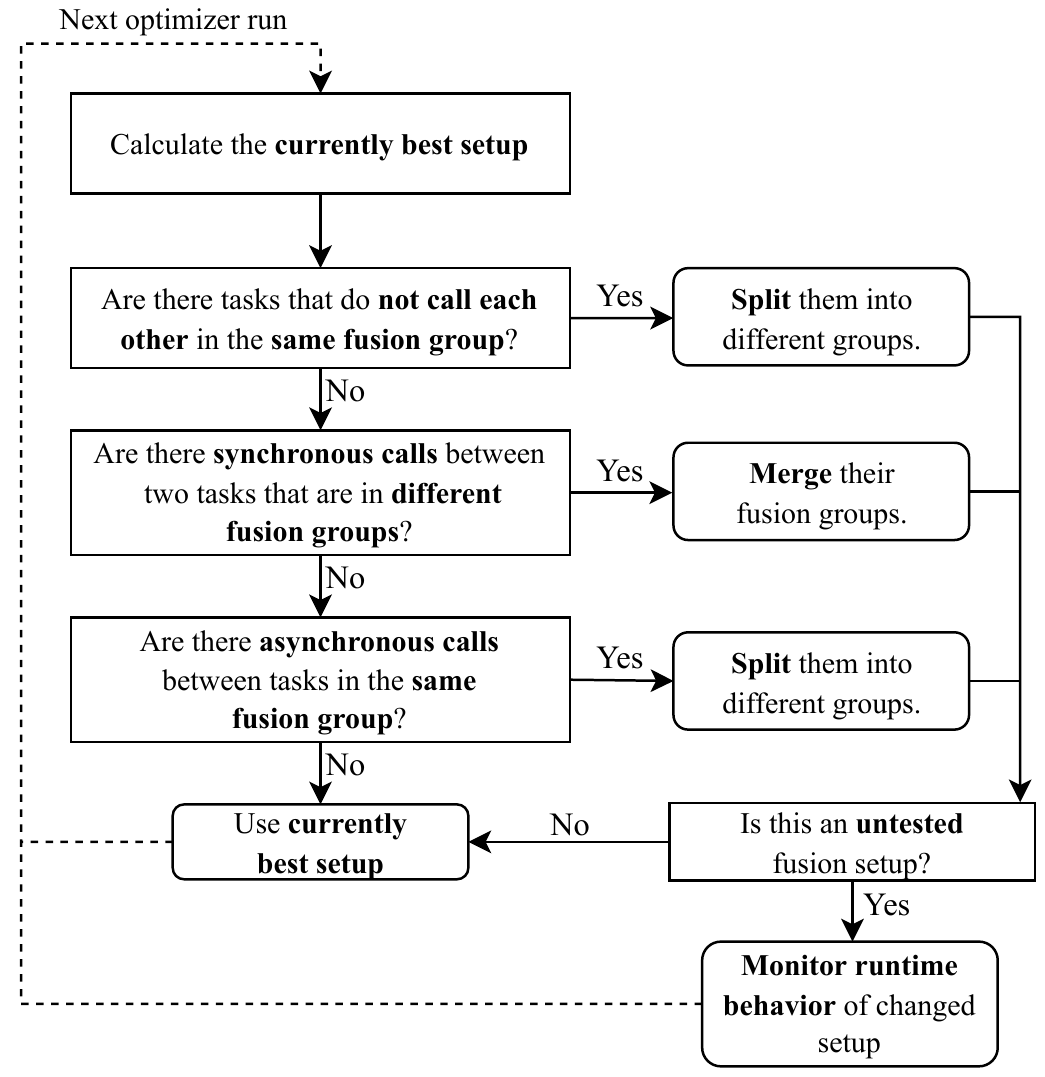}
    \caption{Our heuristic calculates the setup that currently performs best and tries to improve it in three ways: If there are asynchronous or no calls between tasks \emph{in the same} fusion group, move one task into a different fusion group. If there are synchronous calls between two tasks in \emph{different} fusion groups, merge those fusion groups. If a new setup has not previously been deployed, it is considered for the next iteration. Alternatively, the best known setup is used.
    Note that some merges might lead to underprovisioning of resources, which would lead to failed calls. This is not yet handled by our simple heuristic.
    }

    \label{fig:optimizationAlgo}
\end{figure}

Bringing all these aspects together, we propose the heuristic shown in \cref{fig:optimizationAlgo}.
Please note that the Optimizer relies on monitored invocation patterns and not on some kind of composition definition.
This means that whenever the invocation patterns change, e.g., because an application has temporal dependencies or a new version has been released, the heuristic will iteratively adapt the fusion setup.
Additionally,  fusing two existing groups might lead to a function whose requirements exceed those provisioned by the platform.
This can only be detected during runtime and will lead to the splitting of these fusion groups in the next run of the optimizer.

\section{Evaluation}
\label{sec:evaluation}

\begin{figure*}
    \centering
    \includegraphics[width=\textwidth]{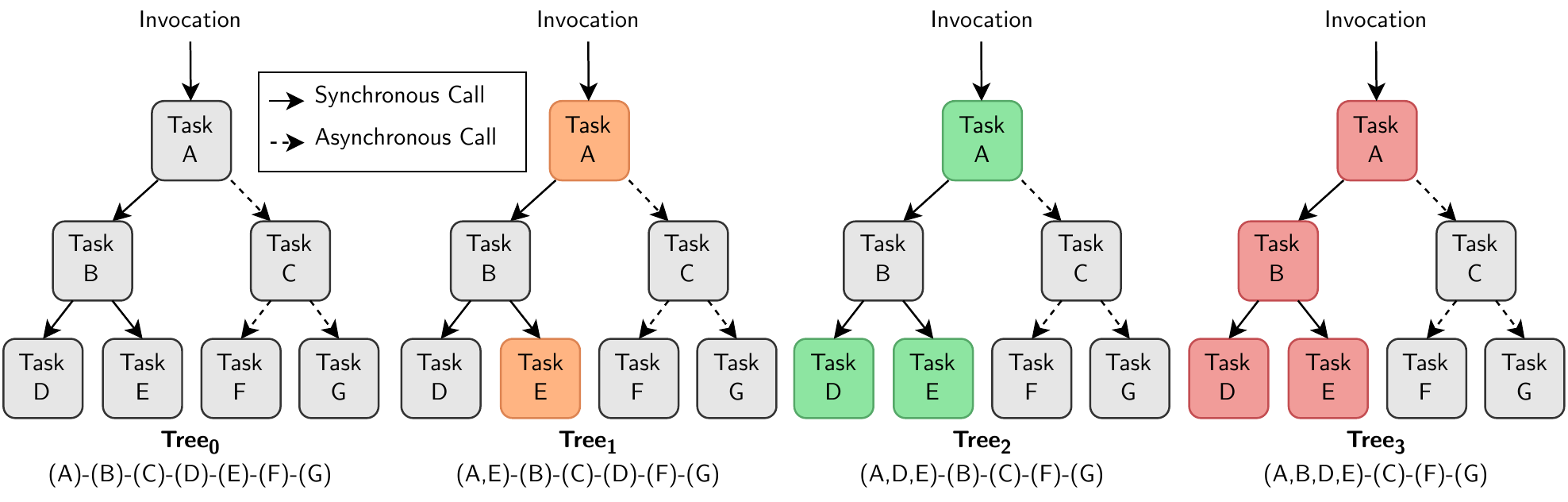}
    \caption{Call graph for the tree experiment with the progression steps the optimizer takes: All non-leaf task call two tasks, with one side of the call tree comprising synchronous tasks and the other side comprising asynchronous tasks. The optimizer starts with the initial setup \textsfcmu{Tree}\textsubscript{0} and changes the fusion group of one tasks until it arrives at \textsfcmu{Tree}\textsubscript{3}, which is the final state for our heuristic where all synchronous invocations are fused together. For this figure, all colored tasks are fused into one fusion group, and all other tasks are in their own fusion group.}
    \label{fig:eval:allInOne}
\end{figure*}

To evaluate our approach, we show that \name{} can be implemented in practice using a proof-of-concept prototype for AWS Lambda (\cref{subsec:implementation}).
Using our prototype, we explore the effects of function fusion with \name{} through two example applications.
The first application is a sample use case built to demonstrate all features of \name{} (\cref{subsec:tree}).
It consists of multiple tasks that are either quick and called synchronously, or are computationally intensive and called asynchronously.
The second case-study is a more realistic implementation of an IoT application, where input from different sensors is used to analyze traffic and stored in a serverless cloud database (\cref{subsec:iot}).
We make our artifacts and prototype available as open-source\footnote{\url{https://github.com/umbrellerde/functionfusion}}.

\subsection{Prototype Implementation}
\label{subsec:implementation}

We implement a prototype of \name{} for AWS Lambda.
We focus on Node.js for this proof-of-concept as it is the most widely-used runtime on AWS Lambda~\cite{Eismann_2021_Review} and an interpreted language, which simplifies dynamic loading of code.
Please note, however, that our approach is extensible for other programming languages and FaaS platforms.

Internal task calls are implemented with an embedded handler that routes the call internally for fused tasks or externally over HTTPS for tasks that are deployed in a different fusion group.
This requires only small changes to the existing programming interface.
In our prototype, the Optimizer is implemented as two Lambda functions:
The first periodically reads function logs from AWS CloudWatch and writes them to AWS S3.
The other uses this information to update the fusion setup.

In our implementation, we chose not to redeploy the entire application whenever the fusion setup is changed.
Instead, each function executable contains the source code for all tasks and updating the fusion setup only ``rewires'' the assignment of tasks to functions by updating an environment variable.
This reduces the time required to deploy new fusion setups, but could be infeasible for some applications with large dependencies, e.g., large machine learning models.
We have also implemented an extended prototype which generates distinct serverless function deployment packages and compared it to the first implementation (see \cref{subsec:discussion:eval})

Due to our implementation of handling remote requests, the first remote request of a cold function often took $\geq$1s.
All following requests were handled within $\leq$50ms.
This limitation is implementation specific and could be sped up by saving this value in a location that is faster to access, e.g. a managed database service, object storage, or the function image.
It could also be possible to use cold start reduction approaches to minimize the impact of these API calls~\cite{paper_bermbach_faas_coldstarts}.

\subsection{Sample Tree Application}
\label{subsec:tree}

In the sample tree application, some computationally light tasks are called synchronously, while other computationally heavy tasks are called asynchronously.
These tasks are called in a tree-like pattern, where one side of the tree consists of synchronous calls, and the other side consists of asynchronous calls.
In the call graph shown in \cref{fig:eval:allInOne}, we mark all fusion setups that were tested by \name{} during our experiments.

In the initial fusion setup in all of our experiments, all tasks are in their own fusion group.
This is the way that functions would be deployed in a normal serverless system to maximize flexibility and reusability.
For all experiments presented here, we configured all functions to use 128MB of memory.

\subsubsection*{\textbf{Request Response Latency}}
\label{sec:eval:split:full}

\begin{figure*}
    \centering
    \includegraphics[width=\textwidth]{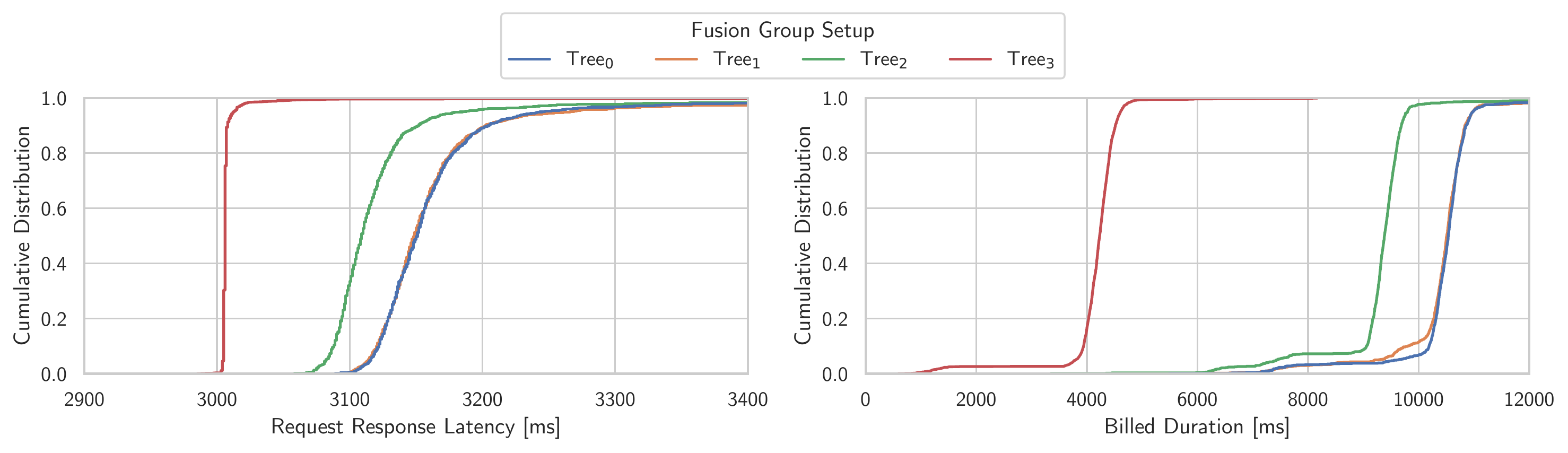}
    \caption{In the tree application with 1,000 invocations per fusion setup, \name{} iteratively improves fusion setup performance: While \textsfcmu{Tree}\textsubscript{0} and \textsfcmu{Tree}\textsubscript{1} perform nearly the same, \textsfcmu{Tree}\textsubscript{3} is an average of 100ms faster than the next best performing setup while also on average incurring less than half the cost of any other tested setup. The final setup has a mean billed duration (i.e., the total runtime of all fusion functions) of 4.1s while all other setups have an average between 9.2s and 10.5s}
    \label{fig:eval:split:full}
\end{figure*}

\begin{figure*}
    \centering
    \includegraphics[width=\textwidth]{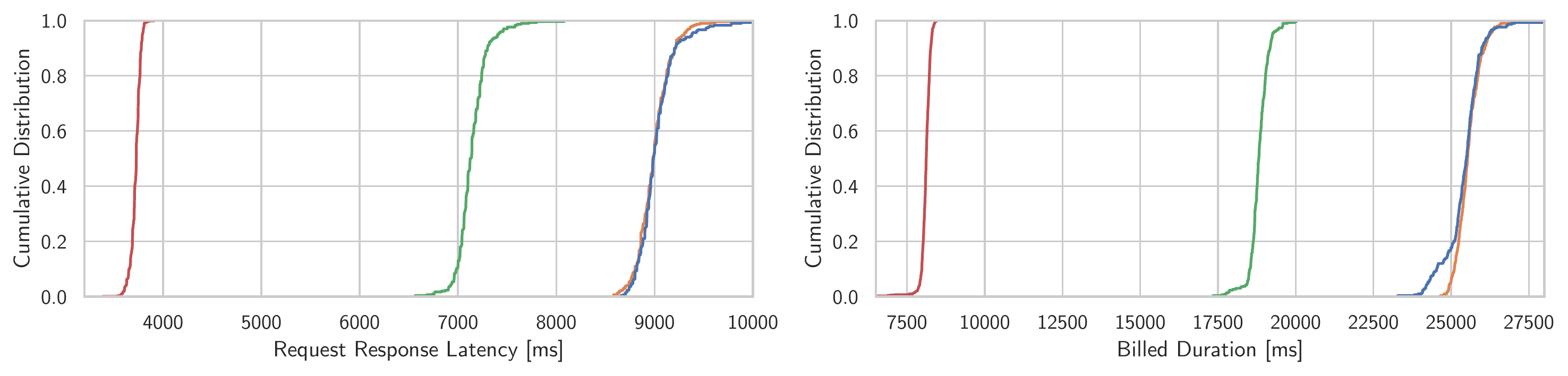}
    \caption{When initiating 300 cold starts for different fusion setups, the fastest setup significantly improves application performance: Almost all invocations are faster and cheaper than any invocation of the next-slowest fusion setup, as this setup avoid cascading cold starts in the application by merging tasks with synchronous calls. The fastest setup has an $rr_{avg}$ of 3.7s, which is 2.4 times faster than the initial setup (9.2s) and 1.9 times faster than the next-best setup (7.1s). The average billed duration starts at 25.4s and goes down to 8.1s (313\% decrease).}
    \label{fig:eval:split:cold}
\end{figure*}

The focus of this evaluation lies on demonstrating that \name{} can be used to optimize the request response latency of invocations.
To test this in a normal use case, we put our prototype under a load of one request per second (rps) for 1,000 seconds.
This low request rate reflects that of common FaaS workloads~\cite{Shahrad_2020} and leads to around three active function instances per function.
After 1,000 requests, we run the Optimizer with the goal of reducing the median request response latency ($rr$) until it can find no further fusion setups to test.
We then calculate median ($rr_{med}$) and average request response latencies ($rr_{avg}$) of invocations, sum up the billed duration of all functions that were triggered by an invocation, and plot their empirical cumulative distribution functions.

The results of the full experiment are shown in \cref{fig:eval:split:full}.
The Optimizer first changed the initial setup
\textsfcmu{Tree}\textsubscript{0} to \textsfcmu{Tree}\textsubscript{1} by fusing tasks \texttt{A} and \texttt{E} together. In \textsfcmu{Tree}\textsubscript{2}, task \texttt{D} is also put in this group.
For the final fusion setup \textsfcmu{Tree}\textsubscript{3}, task \texttt{B} is also added to this group leading to the final fusion setup \fancytt{(A,B,D,E)-(C)-(F)-(G)}.
In \textsfcmu{Tree}\textsubscript{3}, the whole synchronous call tree is in the same fusion group.
Applying the heuristic reduced $rr_{med}$ from 3.1s to 3.0s and reduced the average billed duration per invocation from 10.5s to 4.1s.
This experiment shows that by using function fusion, the median and average latency can be decreased while also reducing cost in a use case where some computationally complex asynchronous functions can be extracted into their own fusion groups.

\subsubsection*{\textbf{Cold Start Request Response Latency}}

We also run experiments to evaluate how \name{} can be used to minimize the request response latency of cold starts.
In this case, we manually create cold starts by replacing the environment variables of the Lambdas and calculate the median request response latency of cold starts.
We create 300 cold starts for every fusion setup, after which the Optimizer runs using the same method we described above.
The results of the tree cold start experiments are shown in \cref{fig:eval:split:cold}.
After testing
\textsfcmu{Tree}\textsubscript{0} ($rr_{med}$=9,206ms, $rr_{avg}$=8,988ms), the optimizer merges the tasks \texttt{A} and \texttt{E} (\textsfcmu{Tree}\textsubscript{1}).
After also merging \texttt{D} into this fusion group (\textsfcmu{Tree}\textsubscript{2}), the Optimizer finished at \textsfcmu{Tree}\textsubscript{3} \fancytt{(A,B,D,E)-(C)-(F)-(G)} ($rr_{med}$=3,721ms, $rr_{avg}$=3,728ms), which is the same setup found in the full experiment.
The order the fusion groups were tested in was the same as in the normal experiment, since the call graph is the same.
The average billed duration of an invocation decreased from 25.4s to 8.1s.

\subsubsection*{\textbf{Discussion}}

Overall, these results show that \name{} can successfully decrease the request response latency of normal invocations as well as cold start invocations in an artificial application.
The fastest fusion setup we tested was the same for both use cases, which indicates that the heuristic we describe in \cref{sec:heuristics} can yield a noticeable performance increase over the initial setup.

\subsection{Realistic IoT Application}
\label{subsec:iot}

\begin{figure*}
    \centering
    \includegraphics[width=\textwidth]{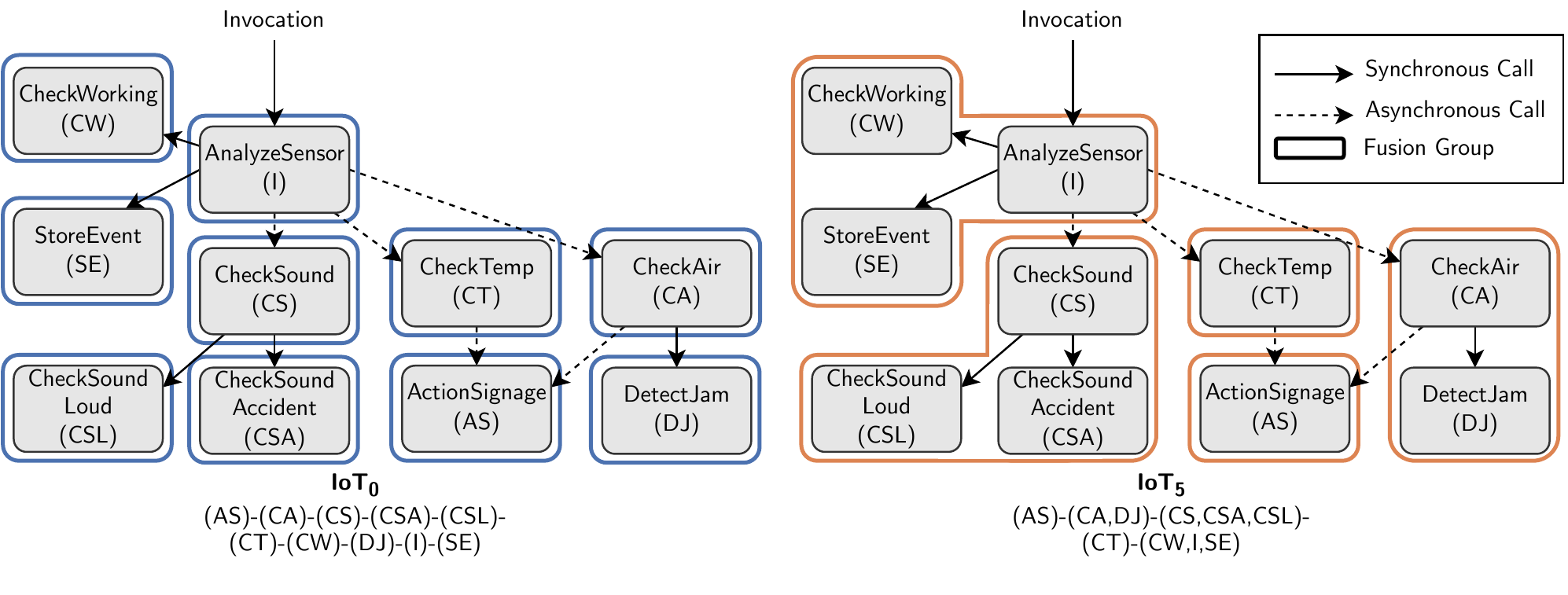}
    \caption{Call graph for the IoT application with the initial (\textsfcmu{IoT}\textsubscript{0}) and final (\textsfcmu{IoT}\textsubscript{5}) fusion setup marked. In the fastest setup, all synchronous call chains are in the same fusion group. Due to space reasons we have refrained from visualizing the intermediate fusion setups \textsfcmu{IoT}\textsubscript{1}-\textsfcmu{IoT}\textsubscript{4} in this paper.}
    \label{fig:eval:iot}
\end{figure*}

To evaluate \name{} in a realistic use case, we present an IoT application, which is a common use case for serverless computing~\cite{Eismann_2021_3c, Castro_2019}.
In this application, sensors placed along roads generate temperature, volume, and air quality readings.
These readings are then analyzed by different tasks and stored in an AWS DynamoDB table.
An overview of the call graph of this application can be found in \cref{fig:eval:iot}.
As is the case in the tree experiment, all asynchronous tasks calculate primes to emulate complex functions.
Additionally, the tasks \texttt{AS}, \texttt{CSA}, \texttt{DJ}, and \texttt{SE} write information into DynamoDB, and the task \texttt{CSL} sends two queries to DynamoDB before writing data itself.
The goal of this use case is to test whether \name{} can optimize more complex fusion setups.
For example, one task is called by multiple other tasks (\texttt{AS} is called by \texttt{CT} and \texttt{CA}), and there is a chain of calls that are alternating between synchronous and asynchronous calls (\texttt{I} asynchronously calls \texttt{CA}, which synchronously calls \texttt{DJ}).
Otherwise, the experiment setup is the same as in \cref{subsec:tree}.

\subsubsection*{\textbf{Request Response Latency}}

\begin{figure*}
    \centering
    \includegraphics[width=\textwidth]{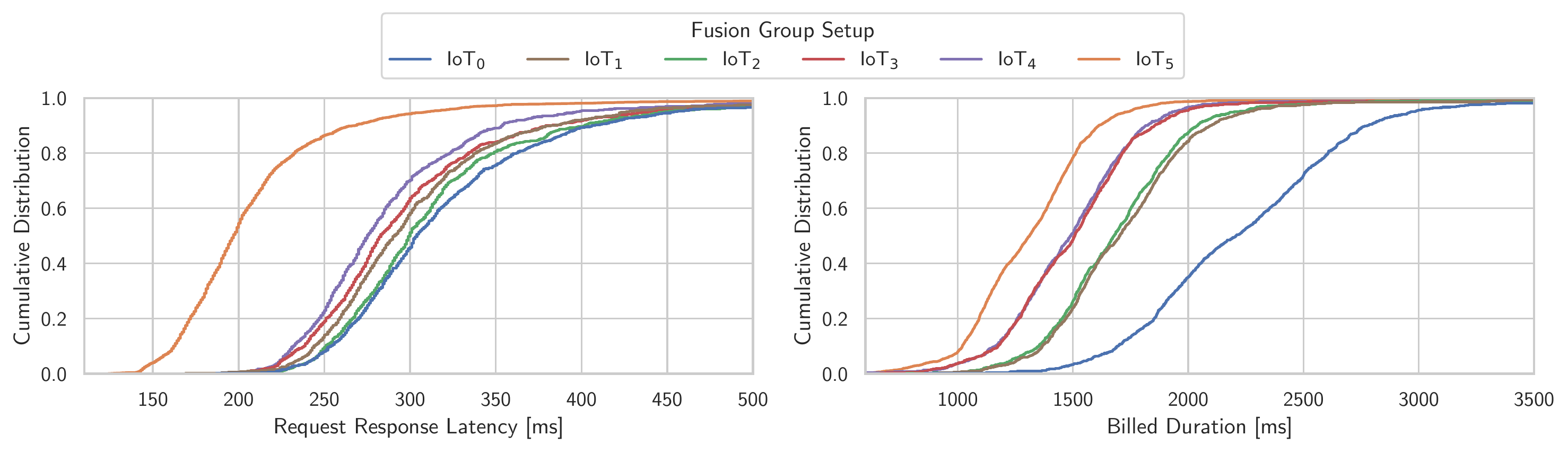}
    \caption{In our test of \name{} on IoT application with 1,000 request per fusion setup, the first four improvements lead to a decreased distribution of request response latency. In the fastest setup, $rr$ is reduced by a further 70\%. While faster fusion setups also incur less cost, this difference is not as noticeable as in the tree application.}
    \label{fig:eval:iot:full}
\end{figure*}

In the IoT experiment (cf. \cref{fig:eval:iot:full}), the Optimizer tried four fusion setups before arriving at the fastest tested setup \textsfcmu{IoT}\textsubscript{5} (\fancytt{(AS)-(CA,DJ)-(CS,CSA,CSL)-(CT)-(CW,I,SE)}).
In \textsfcmu{IoT}\textsubscript{1}, \texttt{CA} and \texttt{DJ} are fused together. Next, \texttt{CS} and \texttt{CSA} are fused together (\textsfcmu{IoT}\textsubscript{2}). In the next step, the optimizer adds \texttt{CSL} to this fusion group (\textsfcmu{IoT}\textsubscript{3}). Afterwards, \texttt{CW} is merged with \texttt{SE} (\textsfcmu{IoT}\textsubscript{4}) and finally with \texttt{I} (\textsfcmu{IoT}\textsubscript{5}).
In this setup, all synchronous tasks are located in the same fusion group while asynchronous tasks are kept seperately.
The median request response latency of the first four setups drops from 304ms in \textsfcmu{IoT}\textsubscript{1} to 275ms in \textsfcmu{IoT}\textsubscript{4}, while \textsfcmu{IoT}\textsubscript{5} has an $rr_{med}$ of 197ms.
\Cref{fig:eval:iot} shows a visual representation of \textsfcmu{IoT}\textsubscript{1} and \textsfcmu{IoT}\textsubscript{5}
\subsubsection*{\textbf{Cold Start Request Response Latency}}

\begin{figure*}
    \centering
    \includegraphics[width=\textwidth]{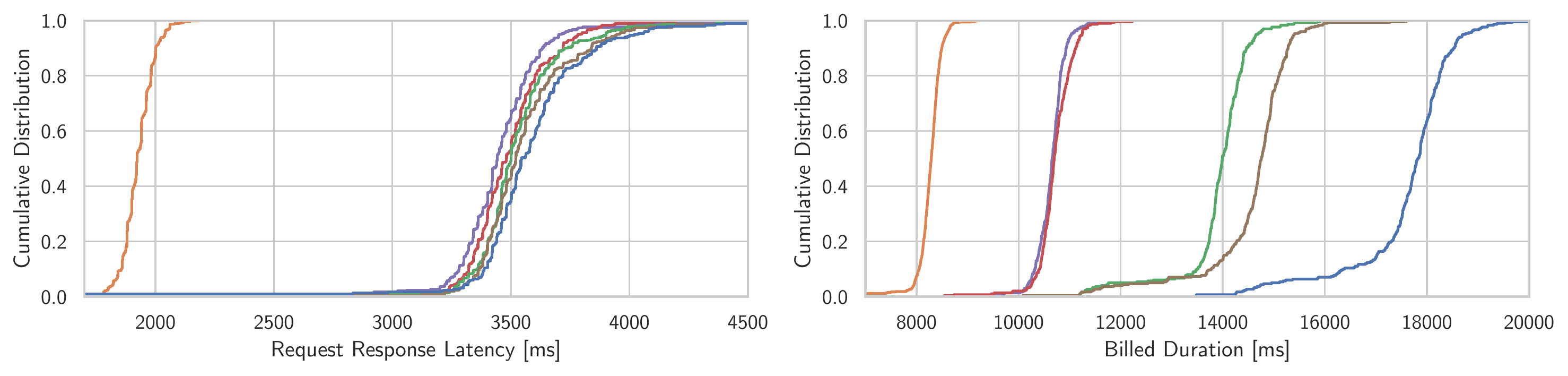}
    \caption{In our test of 300 cold starts per fusion setup, a faster setup can again avoid cascading cold starts, leading to a 53\% (1,000ms) lower $rr_{avg}$. While the fastest setup (\textsfcmu{IoT}\textsubscript{5}) is also the cheapest one, the effects on costs are not as noticeable. Its average billed duration is 78\% of \textsfcmu{IoT}\textsubscript{4} and 47\% of \textsfcmu{IoT}\textsubscript{0}.}
    \label{fig:eval:iot:cold}
\end{figure*}

The results for the cold start experiments are shown in \cref{fig:eval:iot:cold}.
The Optimizer tried the same fusion setups it did in the request response latency experiment and finished with the fusion setup \textsfcmu{IoT}\textsubscript{5}, which has the lowest median request response latency of all tested setups ($rr_{med}$=1,920ms, $rr_{avg}$=1,922ms).
All other setups had a similar distribution of cold start latencies, with a median between 3,423ms and 3,521ms.
This can be explained by the order in which our optimization algorithm tried fusion setups:
Only the fastest setup merges tasks \texttt{CW}, \texttt{I}, and \texttt{SE} into the same fusion group.
In all other cases, task \texttt{I} has to synchronously wait for the completion of \texttt{CW} and \texttt{SE}, which leads to two additional cold starts that happen in the critical path.
While the $rr$ of the slower functions is virtually identical, the total billed duration per invocation has different distributions.
Overall, invocations for the initial fusion group \textsfcmu{IoT}\textsubscript{0} have the highest cost, since this setup creates the most cold starts.
While the fastest fusion setup is also the cheapest one of the tested setups, this shows that request response latency and billed duration do not necessarily correlate.

\subsubsection*{\textbf{Discussion}}

Overall, these results show that \name{} also works in a more complex application that models a real-world setup.
As was the case in \cref{subsec:tree}, the fastest setup was the same for both the normal and the cold start test.
This indicates to us that for many workloads, the heuristic of putting all synchronous invocations into the same fusion group and splitting off all asynchronous invocations is a good starting point from which further optimizations might be possible.
In all cases tested here, the fusion groups tested were always faster than the previous fusion group.
Our simple heuristic might find other local optimums if there are faster fusion setups that are only findable by changing two fusion groups at once.
Since the average billed duration of an invocation in the fastest fusion setup is always lower than the initial setup, we argue that \name{} reduces the total cost of execution.
The prototype also incurs other cost, e.g., the Optimizer function and API Gateway costs, which we argue is not relevant for our evaluation since they are a consequence of our implementation and independent of the architecture.
Other FaaS providers might provide different methods to directly call their Functions that do not incur additional cost, while the number of requests and their runtime is likely to stay consistent across different FaaS providers.

\subsection{Framework Overhead}

\name{} adds a handler to every function that manages calling the different tasks. This adds an overhead for every function and task call.
In an experiment where we called a single empty task once per second for 200 seconds, the handler on average ran for $1.3$ms when the function instance was already warm (standard deviation $\sigma$=1.24), and on average ran for $36.6$ms in cold starts ($\sigma$=23.4).

While calling a task locally has almost no overhead, calling a task remotely requires additional time to send an HTTP request to another function.
When \name{} first calls another function, the Base URL to call other functions needs to be determined.
In our current implementation, this requires two API calls to API Gateway, which take around 1.1s in total.
Consecutive HTTP calls take $\leq50$ms.

The Optimizer adds no additional overhead to function calls, since it runs inside its own function and only reads the CloudWatch logs written by every function call.
Computing the next fusion setup for every 1000 invocations takes around one second, while extracting the invocation data from CloudWatch sometimes takes considerably longer depending on the number of cold starts due to limitations in the CloudWatch API.

\section{Discussion \& Future Work}
\label{sec:discussion}

\name{} abstracts even more operational concerns away from users than classical FaaS applications, since it also automates the task of sizing the functions.
The comparatively simple heuristic we have implemented always found a faster fusion setup than the naive initial state.
Nevertheless, some aspects of our architecture and implication warrant further research.

\subsection{Platform Integration}

Our prototype is implemented on top of AWS Lambda and runs inside the functions.
This creates a performance overhead for every function invocation and limits the information available to the framework.
The same architecture could be implemented as a part of the platform, which could lead to increased performance due to less overhead and additional information that can be used to, e.g., allocate tasks to functions or place related function instances in the same physical machine.
For example, call graph analysis might be useful to preemptively start functions in anticipation of tasks that will need it.
This could also solve platform limitations we encountered during our implementation.
When extracting the call graph information from the logs, the amount of requests that can be sent to CloudWatch to query logs is limited.
The creation of the call graph for cold-start heavy workloads took multiple minutes, which in some cases lead to timeouts of the handling Lambda function.
If this task was managed by the platform, the  call graph could be created directly.
Additionally, the wait time when sending the first remote request could also be decreased by using provider knowledge.
While integration into the FaaS platform could increase performance, our architecture also works when deployed by application developers and can be used until function fusion is supported by platforms.
Because the prototype currently only works for Node.js, the Black Box Principle of the Serverless Trilemma~\cite{Baldini_2017_Trilemma} is violated, i.e., the application source code needs to be in JavaScript and must be available to \name{}.

\subsection{Function Resources}

In our prototype, all functions have access to the same amount of resources.
As, e.g., Akhtar et al.~\cite{Akhtar_2020} and Eismann et al.~\cite{Eismann_2021_Sizeless} have shown, giving different fusion groups access to different amounts of resources might further improve deployment goals, as the amount of available resources directly influences cost per invocation of a serverless function.
A further avenue of research is the addition of hardware accelerators to serverless architectures.
Some FaaS platforms offer optional support for hardware accelerators, such as GPUs that can be used by functions\footnote{\url{https://nuclio.io/}}.
While some tasks can be massively sped up by using functions with access to hardware accelerators, they are also more costly to run.
\name{} could be extended to handle the grouping of tasks to functions with or without hardware accelerators to further optimize applications.

\subsection{Experiments}\label{subsec:discussion:eval}

The use cases for our evaluation are not real-world serverless applications.
We would have preferred to adapt a nontrivial open source serverless application in our experiments, but were not able to find any suitable candidates.
Instead of image recognition, our prototype uses tasks that compute primes.
This makes it easier to test different levels of CPU usage and does not rely on other services (e.g. object storage), which could influence latency.%
In our prototype, all fusion setups use the same deployment package. We also tested whether changing deployment packages to only contain the necessary tasks for their fusion group significantly changes our metrics and found that the average and median latencies as well as costs were within 3\% of each other.
We expect that the impact of changing the deployment unit might be bigger in cases where some tasks have a bigger deployment size than we used in our evaluation, which is around 10kB.

\subsection{Programming Model}

In previous work~\cite{Scheuner_2019}, we have presented an approach using transpiling to fuse tasks in fusion groups, rather than the fusion handler we use in our prototype.
While not an issue in our experiments, this would decrease the deployment package to only the code that is necessary for the function group to run.

In both approaches, the code points that are suitable for function fusion, i.e., task entry points, need to be clearly marked by developers.
Thus, they are mainly intended to be used when developing new applications and not to transform legacy applications into serverless applications.
Spillner et al.~\cite{Spillner_2017} have presented a framework that transforms a Python application into (FaaS) functions, and we may integrate such an approach in \name{} to further abstract operational concerns.

\subsection{Invocation-Dependent Fusion Groups}

In our approach, fusion setups are determined only by information about previous invocations, leading to a performance profile of the application.
These fusion setups are static in the sense that they only change after the Optimizer runs.
Yet it may also be feasible to select a fusion setup based on the type of invocation.
For example, the fusion handler could change its behavior if it detects a cold start, or if the input data adheres to certain properties.
This dynamic behavior would add additional time to the duration of the fusion handler, so more complex computations to determine an optimal fusion setup might be voided by the additional duration this computation adds to the total function duration.

\section{Related Work}
\label{sec:relwork}
In our previous work~\cite{Scheuner_2019}, we have explored the concept of function fusion.
In this vision paper, code is dynamically transpiled for the different fusion groups.
We found that loading the code dynamically during runtime offers more flexibility during prototyping and allows the function code to stay the same, so that switching of fusion groups does not require the code to be re-uploaded.
Elgamal et al.~\cite{Elgamal_2018} present an algorithm that minimizes the cost of functions while keeping the latency in a bound by using function fusion and placing the functions at specific edge locations using AWS Greengrass.
Their approach uses AWS Step Functions, which they identify as major cost driver in their implementation.
In comparison, this paper focuses on AWS Lambda without using edge services, e.g., AWS Greengrass.
In this work, we assume that tasks are supposed to be deployed as serverless cloud functions, yet previous work has also considered the optimal placement of tasks over a varied set of compute services such as Container-as-a-Service platforms~\cite{Czentye_2019}, virtual machines~\cite{Horovitz_2019_VmMlFaaS}, or in a fog environment~\cite{Pfandzelter_2019_FogProcessing,paper_pfandzelter_zero2fog}.

While we target applications composed of multiple tasks in this paper, reducing latency and cost of executing a single serverless function, e.g., by reducing cold starts, has been discussed in multiple previous studies (e.g.,~\cite{Manner_2018_Coldstarts, Bardsley_2018_coldstarts, paper_bermbach_faas_coldstarts}).
Others have used statistical analysis~\cite{Akhtar_2020} or machine learning~\cite{Eismann_2021_Sizeless} to predict optimal function parameters by only looking at some function configurations.
This reduces the number of tests to find optimal function sizes.
The CPU and memory footprint of functions can be reduced by sharing memory between functions running on the same virtual machine~\cite{Mahgoub_2021_SONIC}, by using application-level sandboxing~\cite{Akkus_2018_SAND}, or by using unikernels~\cite{Fingler_2019_Unikernel}.
These ideas are complementary to \name{} and may further improve performance of deployed fusion groups.

Ali et al.~\cite{Ali_2020_Batching} present a method to optimize cost and minimize latency by batching multiple invocations into a single function execution.
Such an approach especially useful if cold starts account for a significant part of the application duration, e.g., if the function needs to download big datasets during startup.
This requires queuing requests, whereas applications optimized by \name{} can process events immediately.

\section{Conclusion}
\label{sec:conclusion}

In this paper we have presented \name{}, an approach to optimize FaaS deployments by combining tasks in fusion groups.
Developers only need to write the application tasks following a familiar function programming model, while our framework automatically fuses different parts of the application into functions and manages their interactions.
By abstracting the invocation of subsequent tasks, this framework can handle calls locally or remotely in another function.
Leveraging monitoring data, \name{} optimizes the distribution of application parts to functions to incrementally optimize deployment goals such as end-to-end latency and cost.

Further, we have presented a heuristic for optimizing FaaS functions based on call behavior of the application.
Using a proof-of-concept prototype of \name{} for the Node.js runtime on AWS Lambda, we have shown that our heuristic can improve request response latency and cost in both a sample application and an IoT use case.
In future work, we hope to present further optimization algorithms using this prototype and to integrate this functionality at a platform level.

\section*{Acknowledgments}

We thank Jun-Zhe Lai who supported this work in the scope of a master's thesis.

\balance

\bibliographystyle{IEEEtran}
\bibliography{bibliography}

\end{document}